# A Comparison of Fundamental Methods for Iso-surface Extraction

JAN PATERA[1], VÁCLAV SKALA[2]

Department of Computer Science and Engineering

Faculty of Applied Sciences, University of West Bohemia

Univerzitní 22, Plzeň

CZECH REPUBLIC

hopatera@kiv.zcu.cz, skala@kiv.zcu.cz

http://herakles.zcu.cz

*Abstract:* In this paper four fundamental methods for an iso-surface extraction are compared, based on cell decomposition to tetrahedra. The methods are compared both on mathematically generated data sets as well as on real data sets. The comparison using mathematical data is made from different points of view such as area approximation, volume approximation. On the other hand, the Hausdorff distance and root mean square are used to compare methods on real data sets. The presented comparison can be helpful when deciding among tested methods which one to choose, as well as when we need to compare a newly developed method with other existing approaches.

## 1 Introduction

In the recent period of time volume data have started to play a significant role in many scientific areas and are spread across many professions. In medical field, various devices, such as Computed Tomography (CT) scanners, Magnetic Resonance Imaging (MRI) scanners produce volume data. The volume data are also produced as a result of mathematical or physical simulations and experiments and researchers need to visualize such data.

---

[1] Supported by the Ministry of Education of Czech Republic; project number MSM 235200005 (Information Systems and Technologies)

[2] Supported by project NoE – 3DTV PLT 511568

There are two main techniques for the volume data visualization. The first approach is based on volume rendering (ray-tracing-like methods), the second one on surface rendering (iso-surface-extraction-like methods). The volume rendering methods are complex and work with the whole volume data. This paper is concentrated on surface rendering methods that visualizes surfaces stored in the volume data (so called iso-surfaces). The extracted iso-surface is determined by a threshold value. All the points on the iso-surface have their value equal to the threshold.

The field of the iso-surface extraction is quite large. There are many approaches used for the iso-surface extraction such as view-dependent techniques, parallel or distributed approaches, external memory (or sometimes called I/O) techniques, multiresolution (LOD) based extractions and others. In general, we can describe the iso-surface generation and visualization process with the following steps:

1. Search for all active cells (cells that are intersected by the iso-surface)
2. The iso-surface and normal vectors approximation within these cells (e.g. by a triangle set)
3. Iso-surfaces visualization (visualization of a set of triangles; different iso-surfaces can be visualized with different colors depending on a selected threshold value, alpha blending, etc.)

The first phase of the iso-surface extraction can be accelerated using a wide set of speed up algorithms [7], [9], [10], [11], [17] or [18]. However, we are interested not that much in speed of the extraction process but in properties of the output set of triangles.

As there are many various methods for the iso-surface generation and each such a method generates generally different approximation of a searched iso-surface for a given threshold, there is no way how to compare the resulting iso-surfaces to each other unless we know how the iso-surface should look like. We try to compare generated iso-surfaces produced by different methods.

Such a comparison can be made with respect to the volume data. When we generate the volume data using some mathematical or physical model, we are able to gain some additional information concerning the object that is utilized to make a comparison more informative and objective. As additional information, we assume e.g. possibility to compute area or volume of such an object. For real data sets, when we do not have any additional information concerning the scanned object, we can just use general approaches for comparison, such as Hausdorff distance or root mean square (RMS) distance.

This paper is organized in the following way. At first, compared methods are described. Afterwards, we will explain used approaches for the comparison and how the data are generated. The last two sections are devoted to the error analysis, methods comparisons and conclusion.

# 2 Method Description

## 2.1 Marching Cubes

There are many kinds of volume data. From simulations, we often get unstructured volume data. In the other hand from medical imaging the output data is structured one. We aimed at comparison of iso-surface generation methods that are used for structured data, especially for regular grids. Compared methods do not differ in the kind of used interpolation but only in the way they divide a cell into tetrahedra. The well-known method is Marching Cubes (MC) method that was firstly published by Lorensen and Cline [12].

The input volume data consist of samples organized into a regular 3D Cartesian grid. From such a grid, we can easily obtain a set of cells. The cell has in this case a cube shape and consists of eight corresponding samples from two adjacent sample planes. Four samples are from the first plane and four samples are from the second plane. MC method processes sequentially all the cells that can be found in volume data. The iso-surface, which we are looking for, is specified by a threshold value.

Each cell is processed separately. Firstly, the cell *index* is computed. The cell has eight vertices, let us name them from A to H, and each vertex has its data value. Depending on a selected threshold the vertex is assigned a binary value *index* = $ABCDEFGH_B$. Each bit of the *index* is 0 when the data value in the corresponding vertex is less than the threshold and 1 otherwise.

Based on the *index*, we are able to distinguish 256 cases how the iso-surface can intersect the cell, because each vertex can be inside or outside of the iso-surface. When the *index* is 0 or 255 the cell is not intersected by the iso-surface, otherwise such a cell is called an active cell. The purpose of the *index* will be described later. For an active cell, normal vectors are computed in all its vertices using symmetric or asymmetric difference of data samples.

Each *index* represents a different case how the iso-surface can intersect the cell. All these cases can be tabularized and easily triangulated using linear interpolation. The triangles vertices lay on the cell edges. Note, that triangles vertices are interpolated only on the cell edges, this will not be true for other methods. Maximum of four triangles per the cell is needed to approximate the iso-surface. For each triangle vertex a normal vector is computed from normal vectors in the cell vertices, using linear interpolation as well.

Each cell face is shared by another cell. Due to such sharing, the iso-surface is continuous among adjacent cells. Note that there can be ambiguous faces at which the triangulation proposed by [12] will produce holes. There are few approaches how to avoid the holes. Ambiguous cases can be detected and a special triangulation can be applied [16]. The cells can be divided into tetrahedra and resulting simplices triangulated in a little bit different way as we will describe in the next section. Other approaches are out of the scope of this paper, see [2], [3], [6], [13], [14], [15].

The algorithm complexity of MC method is *O(N)*, where *N* is the number of all cells.

## 2.1 Marching Tetrahedra

Marching Tetrahedra (MT) method is based on the same principle as MC method. The significant difference is that the cube cell is furthermore split into tetrahedra. There are two main splitting schemes. The cell is divided into five tetrahedra (MT5) [8], [15] or the cell is divided into six tetrahedra (MT6) [15]. There are several ways how the cube cell can be divided into five (e.g. Fig. 1) or six tetrahedra (e.g. Fig. 2).

For the five tetrahedra scheme, it is necessary to alternate two different splitting schemes. Otherwise, the continuity of the extracted iso-surface will not be maintained properly.

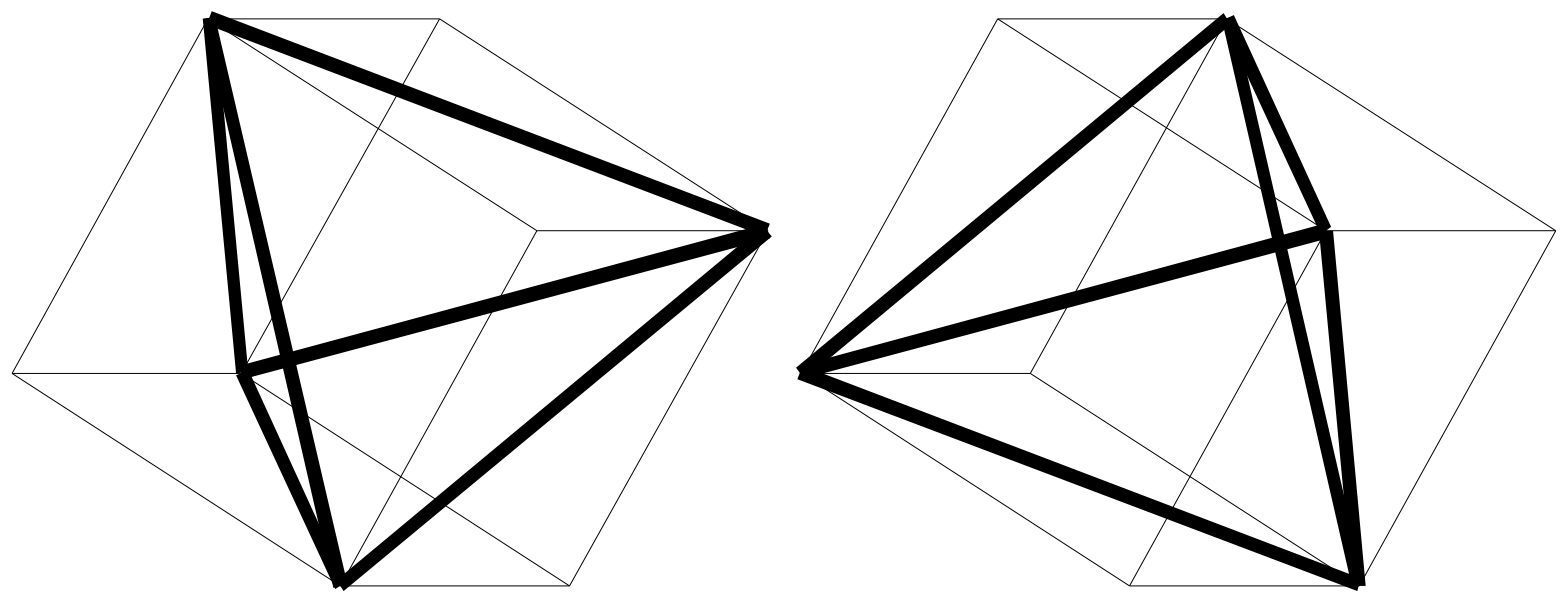

Fig. 1 - MT5 tetrahedra division of the cell

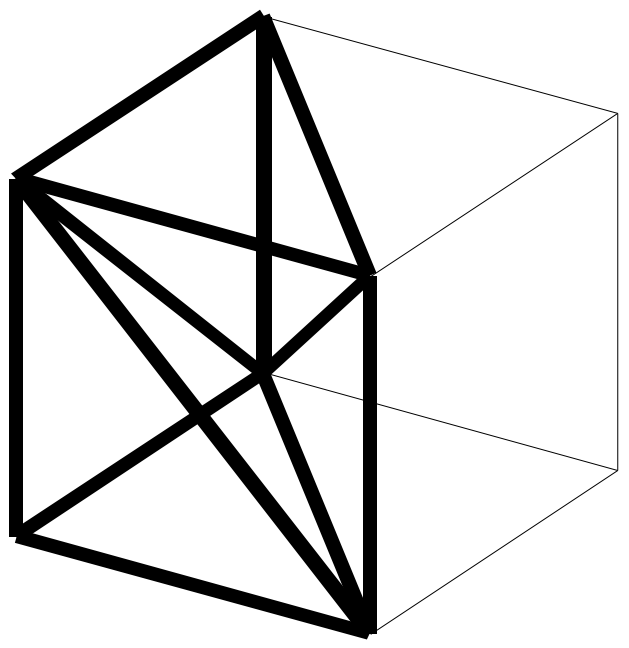

Fig. 2 - Three tetrahedra from a half of the cube, the second half is divided in similar way

After the cell is split into tetrahedra (four vertices), the *index*=$ABCD_B$ for each tetrahedron is computed separately and tetrahedron is processed separately in the similar way as the cube cell in the MC method. There are only 16 possibilities how the tetrahedron can be intersected with iso-surface. These methods generate at most two triangles per tetrahedron.

Five or six tetrahedra decomposition introduces new edges at which the triangles vertices are to be interpolated. For five tetrahedra the interpolation will be held on face diagonals of the cube cell, for six tetrahedra both face and internal diagonals are used.

If we look at five tetrahedra division, there is one tetrahedron with different shape and size. For six tetrahedra splitting, all the tetrahedra are the same.

## 2.3 Centered Cubic Lattice

The last method that will be compared is Centered Cubic Lattice (CCL) method, see [5]. This method is little bit different, because it splits the cube cell into 24 tetrahedra.

The difference is that the resulting tetrahedra are partially shared between adjacent cells and a new data value is introduced to the center of gravity of the processed cell, Fig. 3. There are several ways how to compute the value of the central sample, e.g. the arithmetic mean or weighted mean.

Each tetrahedron is then processed separately in the same way as in MT5 or MT6 methods.

As well as in previous methods this kind of splitting introduces new edges at which the interpolation will be made. These are edges among adjacent central points.

In this division scheme, all the 24 tetrahedra are the same as to the dimensions (similarly to MT6 method).

There are also other possible decompositions of the cube cell, e.g. [19] that decomposes parallelepiped cell into two tetrahedra and one octahedron. These techniques were not included into our study.

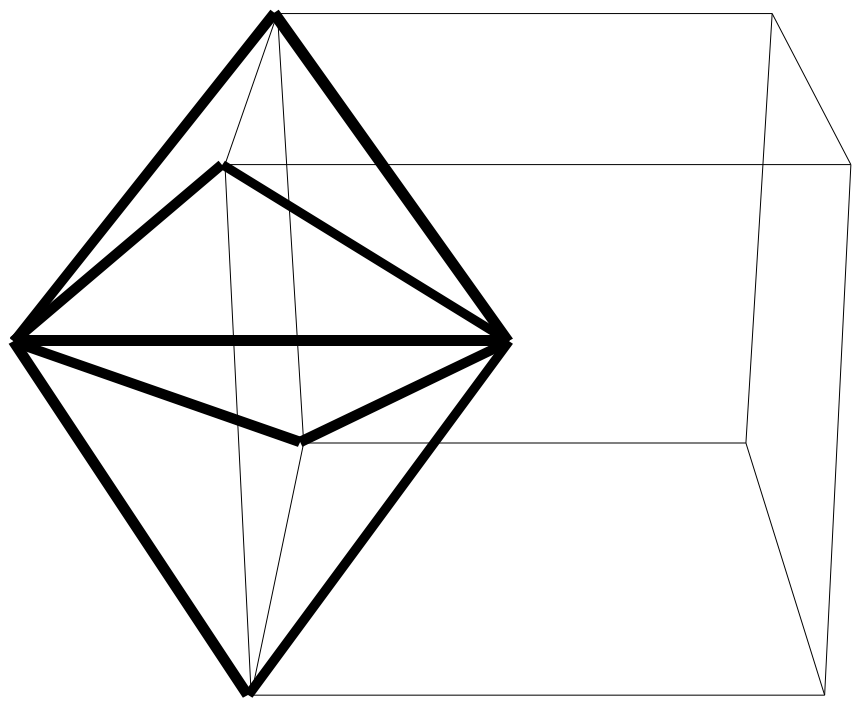

Fig. 3 - Centered Cubic Lattice division for one cell face

# 3 Comparison Approaches

## 3.1 Hausdorff Distance

As mentioned before, we use Hausdorff distance [20] for comparisons mainly for iso-surfaces that are extracted from real data sets. At first, we define a distance between a point *p* (from surface *S*) and a surface *S'* (with points *p'*) as

$$d(p, S')=\min||p-p'||,$$

for all *p'* from *S'*. Now we can define Hausdorff distance between two surfaces *S* and *S'* as

$$d_H(S,S')=\max d(p,S'),$$

for all *p* from *S*. Note important thing that Hausdorff distance is not symmetrical $d(S,S')\neq d(S',S)$. When we call $d(S,S')$ a forward and $d(S',S)$ a backward distance, we can define a symmetrical Hausdorff distance [1] as

$$d_{SH}(S,S')=\max(d(S,S'), d(S',S)).$$

The symmetrical difference provides better error measurement for two surfaces. We utilized a METRO software tool (described in [4]) for accurate computation of Hausdorff distance of two discrete surfaces (triangle meshes). The METRO tool was mainly used to compare original mesh with its simplified (e.g. decimated) version. We use it for comparison of two iso-surfaces, each generated with different method.

## 3.2 Root Mean Square Distance

We use also the Root Mean Square (RMS) of computed distances. RMS distance in discrete case is defined as [20]

$$RMS(S, S') = \frac{\sqrt{x_1^2 + ... + x_n^2}}{n},$$

where $n$ is a number of points of a mesh $S'$, $x_i$ (where $i=1..n$) represents the distance of corresponding point $p_i'$ from $S$ $x_i=d(p_i', S)$. We compare $S'$ to $S$.

Note that RMS is not symmetrical as well as Hausdorff distance. We do not use symmetrical RMS distance in our tests, thus it is not defined here. This measurement is computed with METRO tool as well.

Both the Hausdorff distance and the RMS distance are calculated according to some source mesh using METRO tool. As such a mesh, we use a mesh generated with MC method.

## 3.3 Mathematical Data

At first, we should mention how the testing data are generated from basic mathematical objects. For such objects we need to know an equation. Let us consider for example a sphere. Each vertex of a regular grid has its coordinates and we have to assign it a value. The vertex value is computed as a distance of the grid vertex (known coordinates) from the object surface (known equation). The zero threshold then represents the object surface in volume data.

As we know the object equation and its dimensions, we are able to compute some additional information concerning the object, such as surface area, object volume, triangles position difference from the object surface, etc. We believe that these properties are worth to compute, because they can help us to differentiate among the quality of methods.

**Surface area** – the iso-surface is generated by an extraction method in a form of a set of triangles. We compute the total area as a sum of all triangles area. Than we can compute the area of mathematical object and compare it with iso-surface area obtained. For special objects such as sphere, we are able to track the error behaviour dependency on the sphere radius.

**Volume enclosed with the iso-surface** – for basic objects the volume is computed using appropriate formula. The volume enclosed with the iso-surface is computed in the following manner (for tetrahedra only). There are three cases for a tetrahedron:

1. The whole tetrahedron is outside of the iso-surface – does not affect the total volume computation.
2. The whole tetrahedron is inside – the whole tetrahedron contributes to the total volume. The tetrahedron volume is computed easily.
3. The tetrahedron is intersected with the iso-surface – we have to compute the part of the tetrahedron which is inside of the iso-surface. As there are at most two triangles generated per tetrahedron, these triangles form two small tetrahedra with appropriate tetrahedron vertex and we are able to compute the volume of the tetrahedron part which contributes to the total volume.

**Triangles position difference** – we measure the difference between triangle center of gravity and object surface. This gives us information about triangles position difference compared to the object surface.

The three mentioned geometric properties are the main aspects that we used for extraction methods output comparison. The obtained results are showed in the next section.

# 4 Results

At first, we should describe the data sets used for our comparisons and give the reasons why we chose them. The main part of the used data set is a set of mathematically generated objects, Fig. 4. A real data set was used to show how the Hausdorff distance is dependent on applied iso-surface extraction method. The brief description of used data sets follows in upcoming paragraphs.

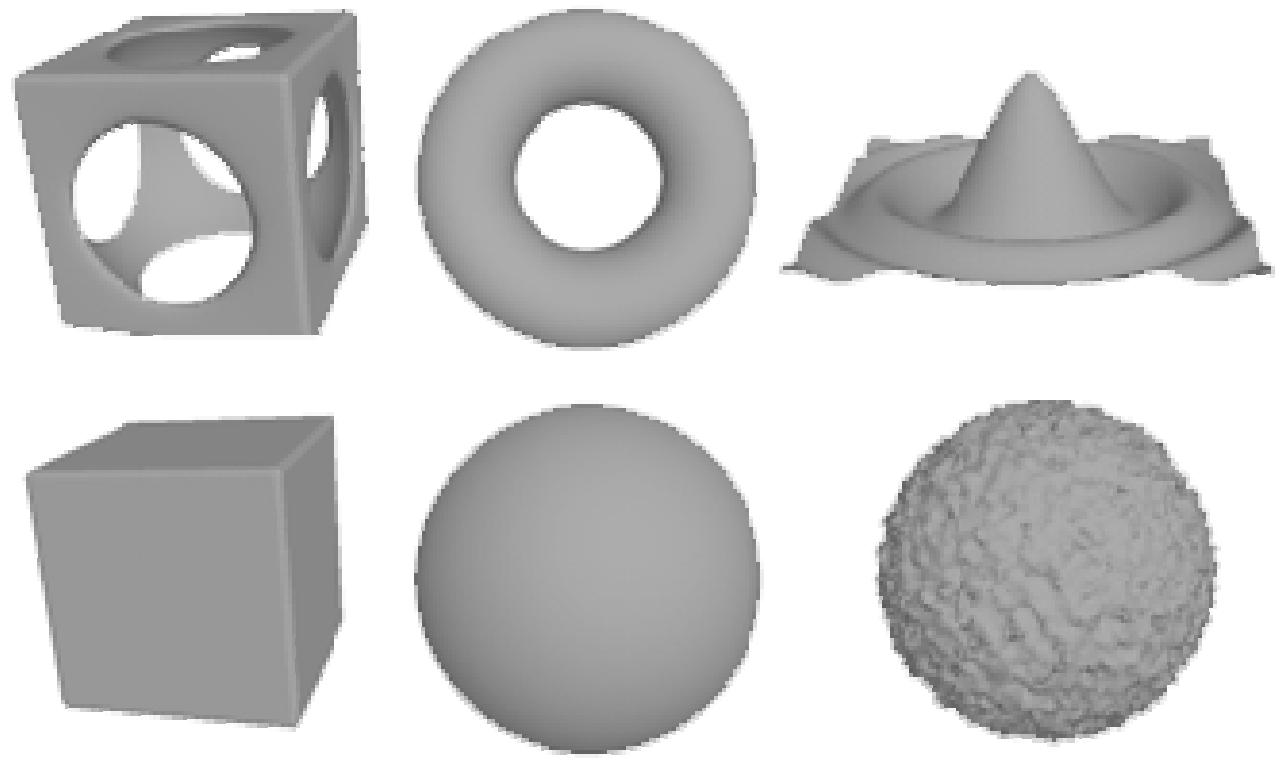

Fig. 4 - Objects (csph, torus, sombrero, cube, sphere and noisedsph)

## 4.1 Used Objects

**Sphere** – sphere is an example of an object that we use to follow the error behaviour dependency on sphere radius. The sphere equation used for data generation is a modified implicit equation

$$F(x,y,z) = \sqrt{(x-s_X)^2 + (y-s_Y)^2 + (z-s_Z)^2} - r$$

where *x*, *y* and *z* are samples coordinates, *sx*, *sy* and *sz* are the sphere centre coordinates, *r* is sphere radius and *F(x,y,z)* is a corresponding sample value. This equation assigns data value to all the volume data samples. The sphere is then represented with a zero threshold iso-surface. The samples that are inside of the sphere have negative value, on the sphere zero value and samples placed out of the sphere have positive value. The sample value represents the distance of the sample from the sphere surface. The radius was 25 in our experiments.

Cell edge has a length 1 for our purposes. The object dimensions (e.g. radius, edge length) are then related to a cell edge length.

**Noised sphere** – (noisedsph) to study the influence of the noise to the shape of the output set of triangles we generate a noised sphere. The random noise is introduced (added) to all samples of the volume data. The size of the noise is given in percentage from the sphere radius size. We used radius 25 and 10% noise.

**Cube** – this kind of an object we use to follow the behaviour and properties of the iso-surface on edges. We will show the iso-surface difference mainly visually. Data are generated similarly as in the previous case using the distance of sample from the closest face, edge or vertex. The inner, on surface and outer samples have the negative, zero and positive value respectively. Cube was generated using *a=b=c=4*2.

**Cube minus sphere** – (csph) such an object was constructed to combine both properties of the sphere (*r=25*) and cube (*a=b=c=42*). The generation of it is a little bit complicated. At first, the cube is generated in the volume data. Afterwards, the values of all samples that are closer to the sphere than to the cube are modified to the new distance.

**Torus** – is the typical mathematically generated object. Torus is defined with the following equation [20]

$$F(x,y,z) = \sqrt{(c-\sqrt{x^2+y^2})^2 + z^2} - a$$

where *x*, *y* and *z* are samples coordinates, *c* is a torus main radius, *a* is a torus secondary radius and *F(x,y,z)* is a corresponding sample value. The samples value are negative, zero or positive as well. Torus dimensions are *c=20* and *a=42* in our case.

**Sombrero** – is the last mathematically generated object we use. It is a surface defined with the mathematical equation (taken from Derive mathematical program)

$$F(x,y,z) = y - \frac{a \cdot \cos(b \cdot (x^2 + z^2))}{c + x^2 + z^2}$$

where *x*, *y* and *z* are sample coordinates and *F(x,y,z)* is a corresponding sample value and *a*, *b* and *c* are constants modifying the shape of the function. Sombrero parameters we used are *a=12, b=0.25* and *c=3.*

**Real data sets** – Samples of real data set have only positive values that represent a density of the space in the sample position (we used engine.vol, ctmayo.vol and hplogo.vol sets).

## 4.2 Tests and Results

For all our mathematically generated objects, we are able to compute the triangles position difference compared to the mathematical object. Firstly, a triangle center of gravity is computed. As we have the routines for point to object distance computation, we can compute the distance of the center of gravity of the triangle from the appropriate object. The overall position difference $P_{ERR}$ is computed as

$$P_{ERR} = \frac{\sum_{i=1}^{n} |\, objDist(\mathbf{O}, \mathbf{T_i}) \,|}{n}$$

where $\mathbf{T}_i$ (*i* goes from *1* to *n*) is the center of gravity of the *i*-th triangle, *n* is the number of triangles and *objDist*(**O, X**) is the distance of point **X** from an object **O** surface.

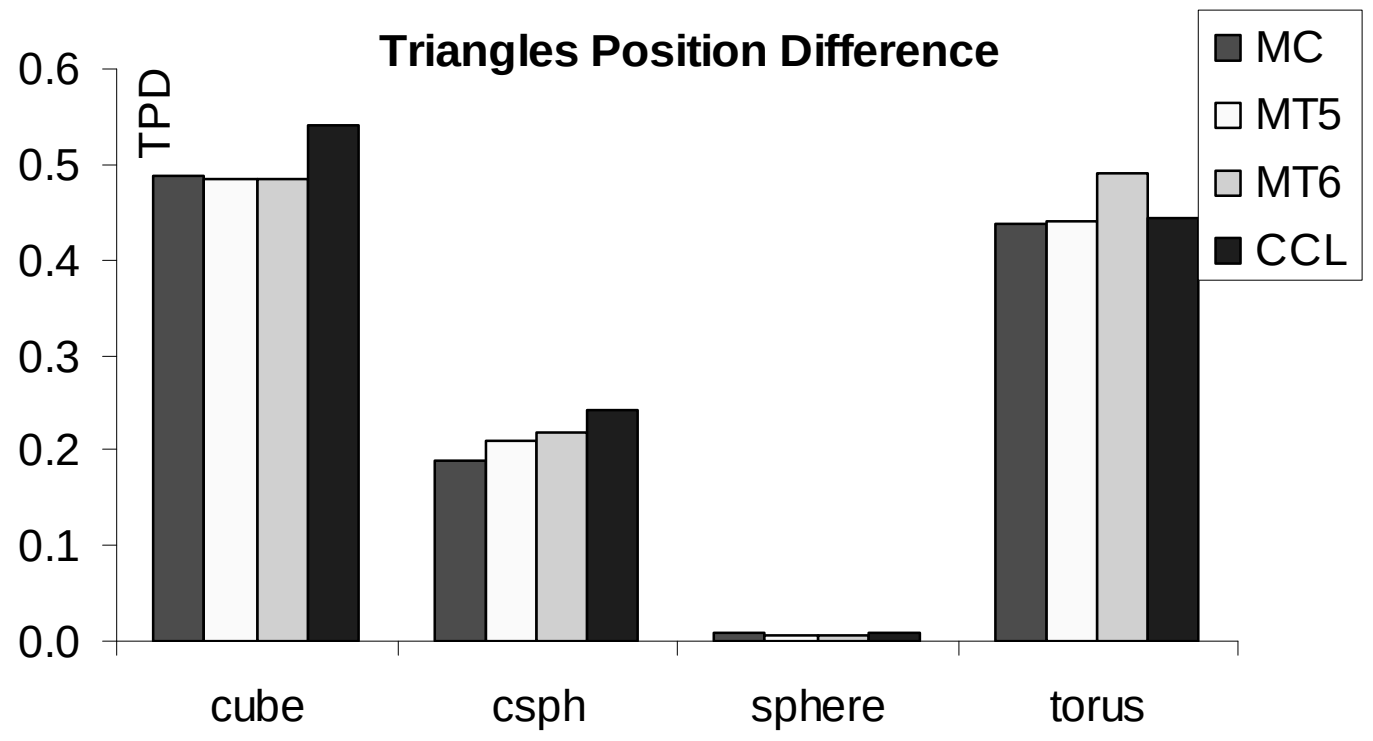

Fig. 5 - Triangles position difference comparison (edge vs. smooth object)

The position difference for a sombrero object was slightly smaller and similar to the results obtained for a sphere. For a cube the CCL method gives the worst results, see Fig. 5. This is probably due to different interpolation of the cube edges (Fig. 6). A csph object has more edges than a cube itself. The more tetrahedra we create the worse results we get. Surprisingly for a torus the MT6 method gives the greatest position difference. We think this is because of the interpolation at a cell interior edge (the longest one).

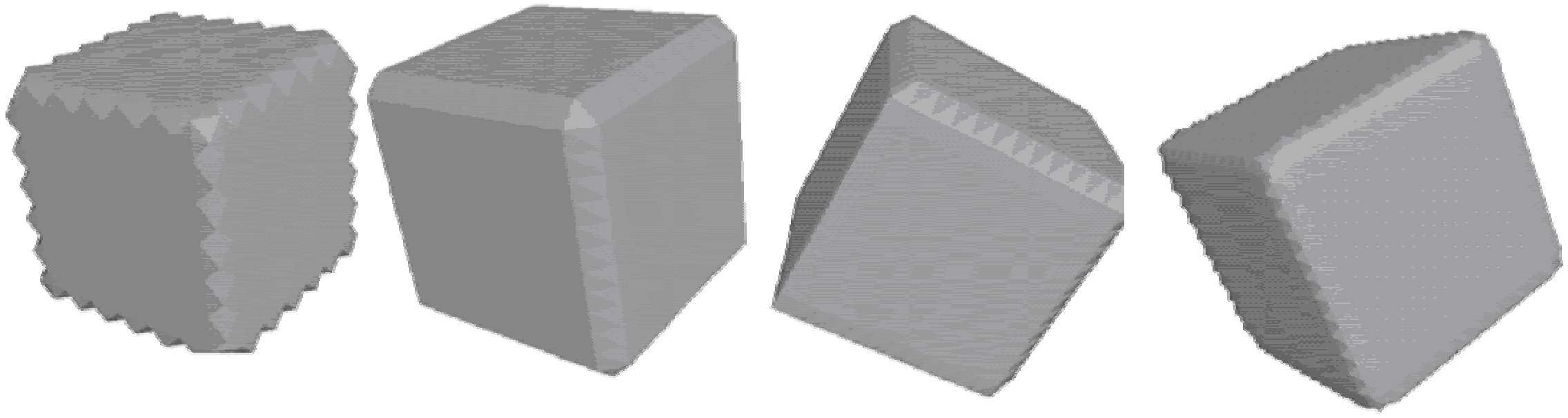

Fig. 6 - Iso-surface on edges (MT5, MC, MT6, CCL)

Note that RMS distance is related to the MC method. For a sphere and a torus the obtained results were slightly less than results for a sombrero. Again, when the object has edges the CCL method is the worst from the view of RMS distance, see Fig. 7. For noisedsph object the CCL method gives the best results. We suppose that the central cell sample value computation (using arithmetic mean) filters data a little bit as well.

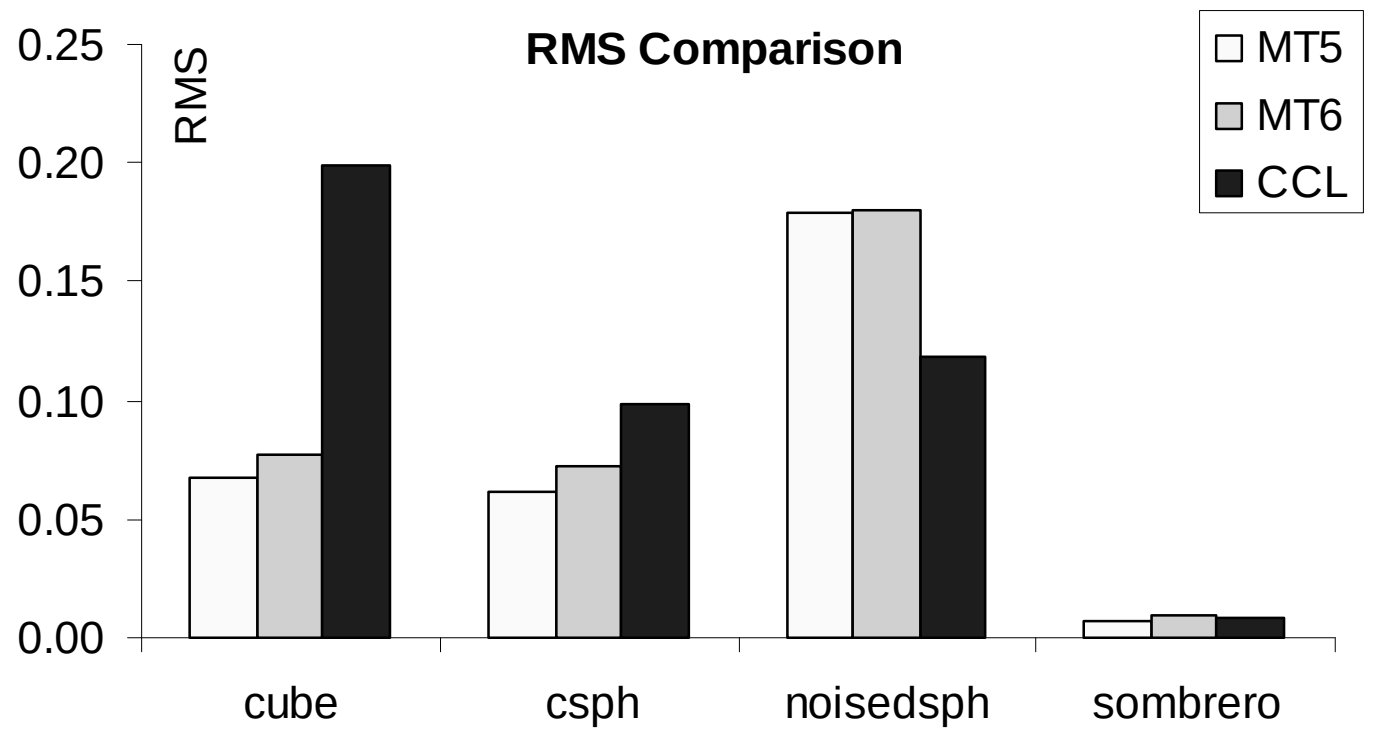

Fig. 7 - RMS distance histogram

Again, a sphere and sombrero give approximately similar results compared to torus. From the view of Hausdorff distance the MT6 method gives the worst results for all tested objects, see Fig. 8. As you can see for noisedsph the CCL method is the best choice. The best choice in this case is probably MT5 method because it does not generate as much triangles as CCL method.

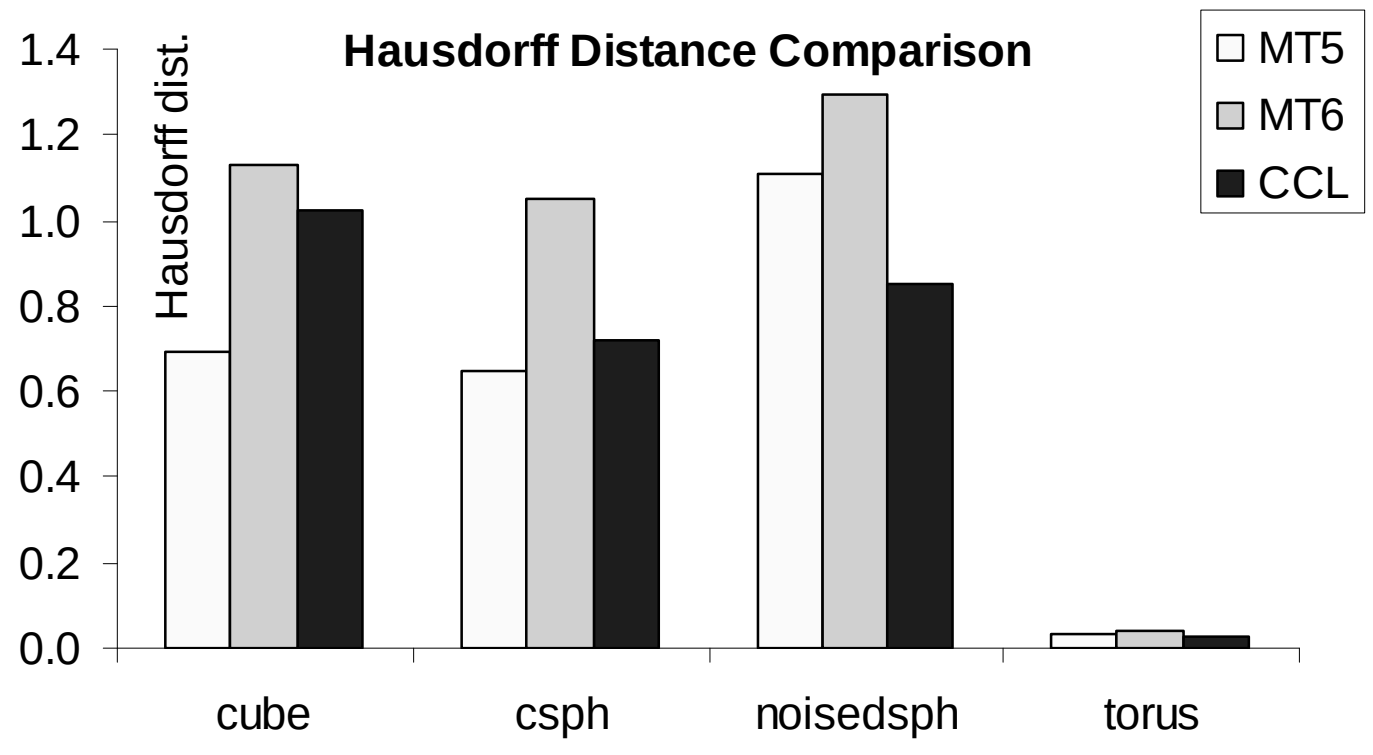

Fig. 8 - Hausdorff distance histogram

The more tetrahedra is used the larger area is extracted for all tested objects that have edges, see Fig. 9. The results in Fig. 9 and Fig. 10 are relative due to mathematical results. For objects like torus (does not have edges) the results were approximately the same as for a sphere. We think that for the area approximation purposes the best choice is MC method.

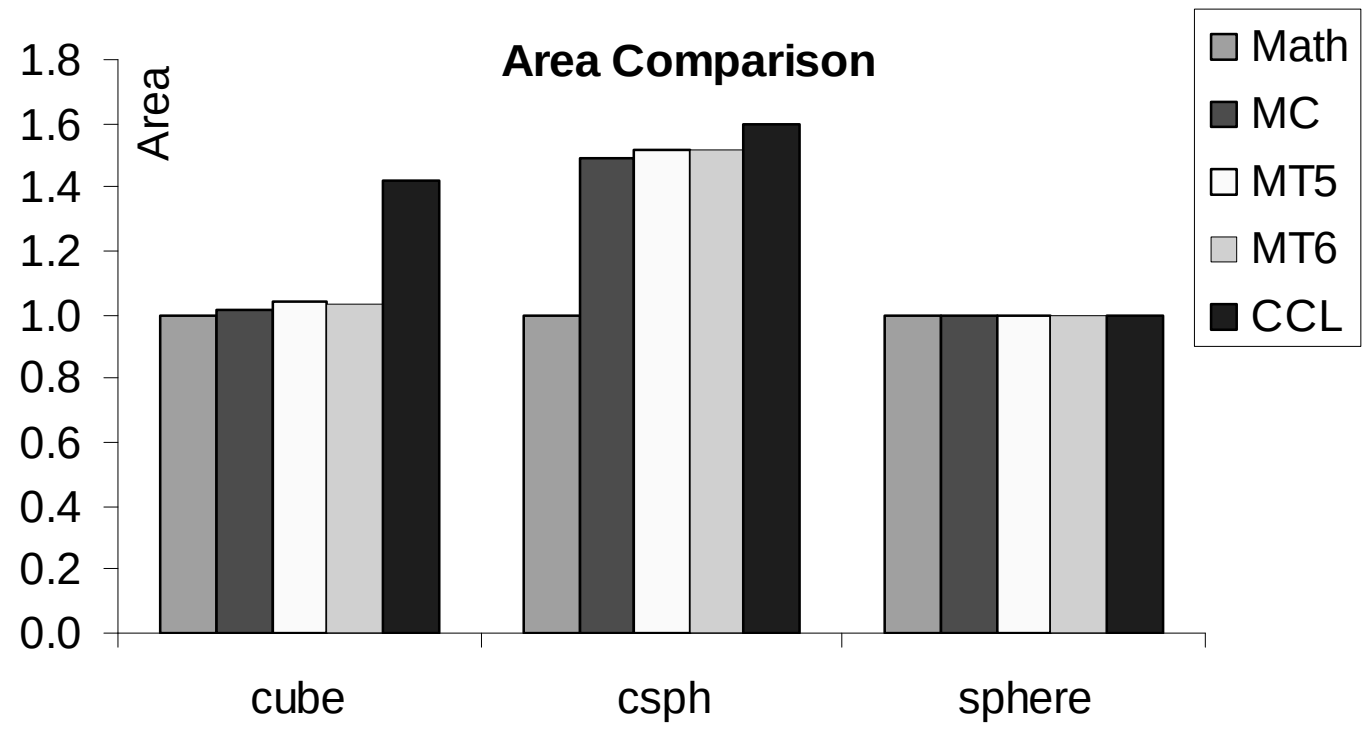

Fig. 9 - Area comparison (relative to mathematical volume)

The MT5 method is in most cases slightly better than MT6 method and both methods are approaching to the original volume from below, see Fig. 10. The CCL method in the other hand is in most cases approaching mathematically computed volume from above. MC method is not included because it is hard to compute the volume enclosed with the iso-surface (due to 256 cases).

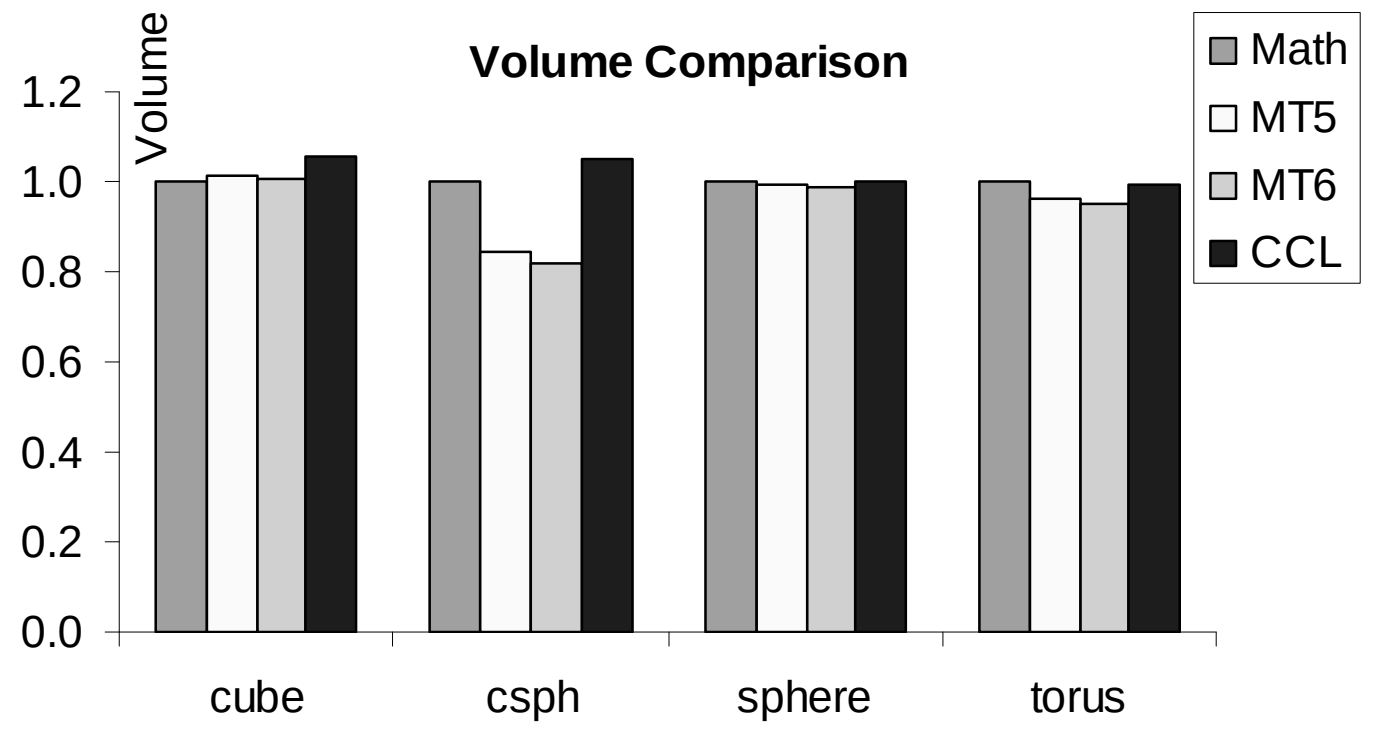

Fig. 10 - Volume comparison (relative to mathematical volume)

## 4.3 Sphere Additional Test

A relative volume error is defined in a following way

$$Error = \frac{V_{TR} - V}{V}$$

where $V_{TR}$ is a volume enclosed with iso-surface triangles, $V$ is mathematically computed volume of the sphere.

The CCL method is the best choice for the volume approximation, see Fig. 11. We assume that it is due to high number of tetrahedra. The CCL method error oscillates about zero value. MT5 gives slightly better results than MT6 method. The progress of error is similar. Both methods are approaching the zero error from below. Another thing we compare is a number of extracted triangles.

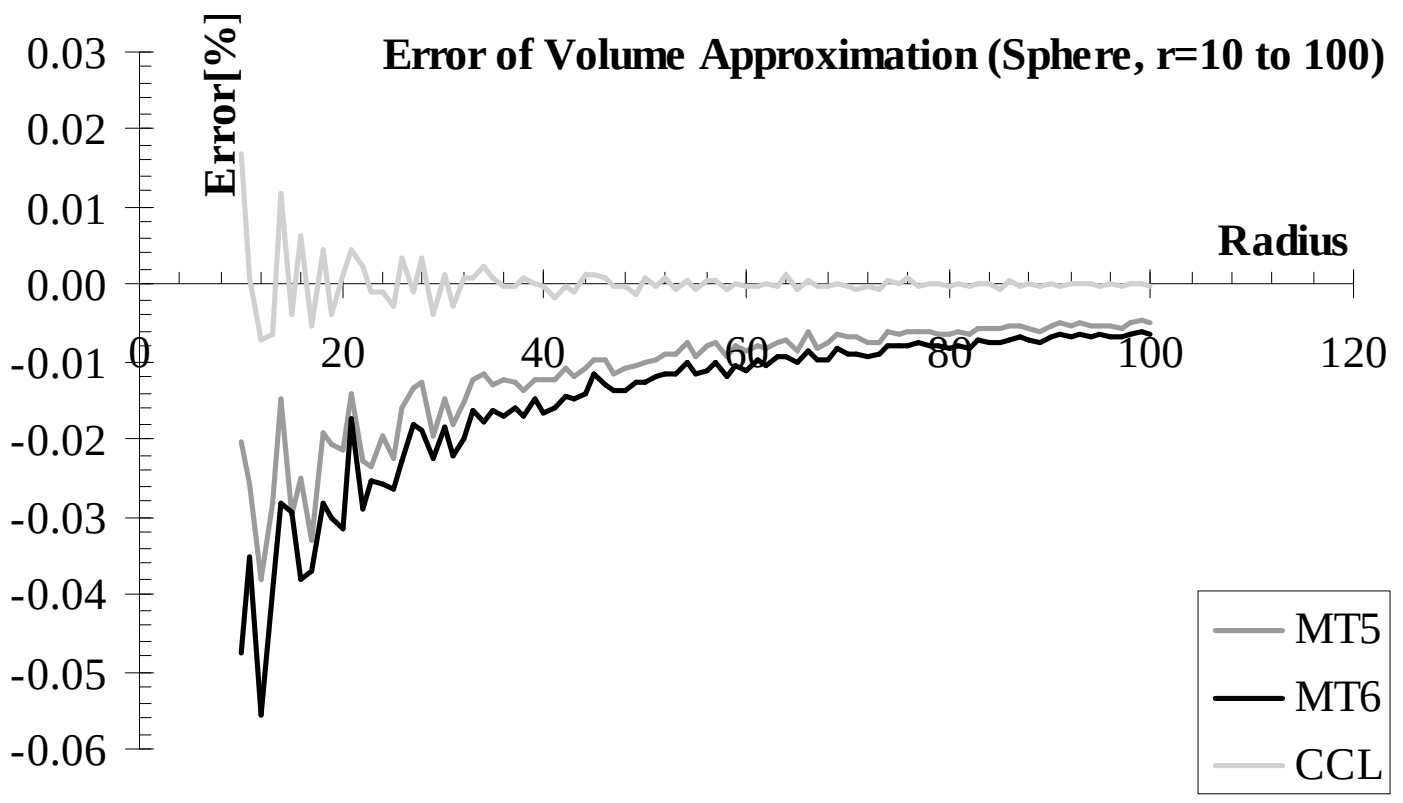

Fig. 11 - Sphere volume error graph

It is a known fact that a number of generated triangles is mainly dependent on the type of the cell division, see Fig. 12. MC works with a cube cell (at most four triangles per cell) and it does not divide it into tetrahedral (at most two triangles per tetrahedron). MT5 divides the cube cell into 5 tetrahedra, MT6 into 6 tetrahedra. In fact, CCL divides the cube cell into 24 tetrahedra, but these tetrahedra also contain parts of adjacent cube cells. When we sum the volume of all 24 tetrahedra, we obtain two cube cells volume, so on average 12 tetrahedra per cube cell.

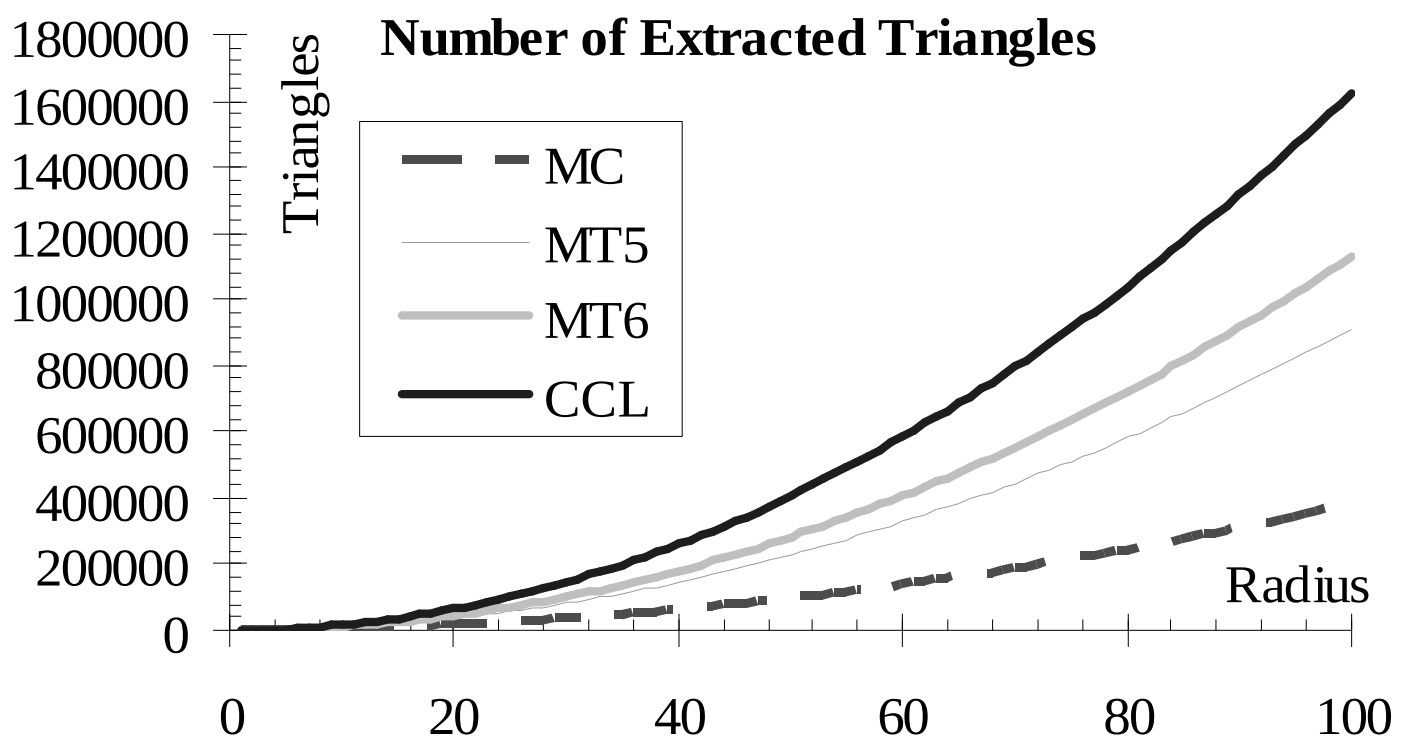

Fig. 12 - Number of extracted triangles

# 4 Conclusions

We compared fundamental methods for the iso-surface extraction evaluating Hausdorff distance, RMS distance, triangles position difference and iso-surface area and volume.

Hausdorff distance is in fact the biggest distance between two compared surfaces (extreme distance). In general, we are more interested in average distance between two surfaces (the RMS distance). In this case, the CCL method generally gives worse results compared to other methods. If we look at a position difference, the MC method seems to be generally the best one. The quality of the extracted set of triangles for noised sphere was in general bad. Interesting is that a volume of objects is approximated with the similar difference no matter of method used except for csph object.

It is important to realize that for real data we do not know the exact area or volume of the object. Hence, the speculations such that the Hausdorff distance is bigger or lower are not completely correct.

## Acknowledgements

We want to thank to Dr. Ivana Kolingerová for her help and support during preparation of this paper.

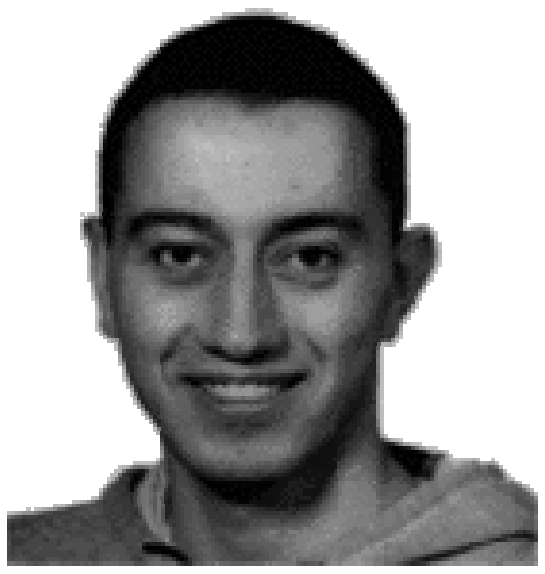

Ing. Jan Patera (http://zcu.cz/~hopatera) is a PhD student and a part-time tutor at the Department of Computer Sciences at the University of West Bohemia in Plzeň in Czech Republic. He graduated in the field of computer graphics at the University of West Bohemia in 2002. He is a member of the Center of Computer Graphics and Data Visualization (CGDV). His research activities concern volume data, iso-surface extraction, algorithms and data visualization.

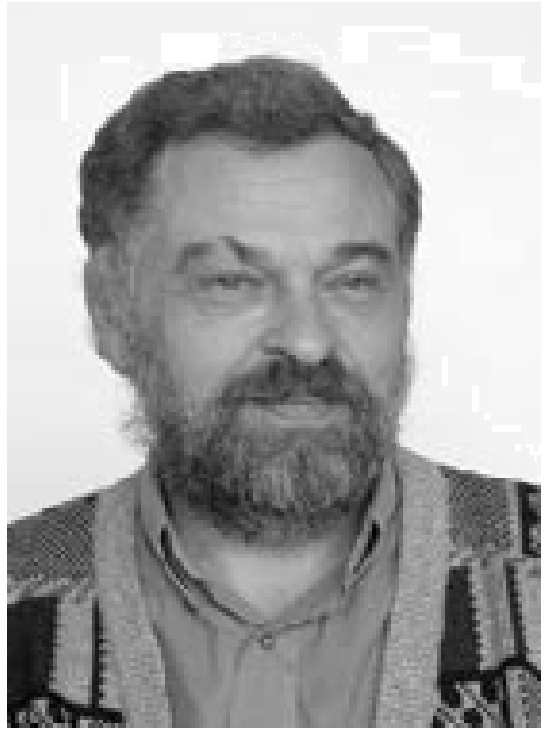

Vaclav Skala is a full professor of Computer Science at the Faculty of Applied Sciences at the University of West Bohemia in Plzen, Czech Republic. He is responsible for courses on Computer Graphics, Algorithms for Computer Graphics, Visualization, Multimedia Systems, Programming in Windows, .NET Technologies at the Department of Computer Science. He is a member of The Visual Computer and Computers&Graphics editorial boards, Eurographics Executive Committee and member of program committees of established international conferences. He has been a research fellow or lecturing at the Brunel University (London, U.K.), Moscow Technical University (Russia), Gavle University (Sweden) and others institutions in Europe. He organizes the WSCG International Conferences in Central Europe on Computer Graphics, Visualization and Computer Vision (http://wscg.zcu.cz) held annually since 1992 and .NET Technologies conferences (http://dotnet.zcu.cz). He is interested in algorithms, data structures, mathematics, computer graphics, computer vision and visualization. He has been responsible for several research projects as well. Currently he is a director of the Center of Computer Graphics and Visualization (http://herakles.zcu.cz).